\documentclass[prb,preprint]{revtex4}
\usepackage[utf8]{inputenc}
\usepackage{graphicx}
\usepackage{bm}

\begin{document}

\title{Electric current control of spin helicity in an itinerant helimagnet}
\author{N. Jiang$^{1}$}
\email{jiangnan@g.ecc.u-tokyo.ac.jp}
\author{Y. Nii$^{2}$}
\author{H. Arisawa$^{2}$}
\author{E. Saitoh$^{2,3,4,5}$}
\author{Y. Onose$^{2}$}
\affiliation{$^{1}$Department of Basic Science, The University of Tokyo, 3-8-1 Komaba, Meguro-ku, Tokyo 153-8902, Japan.\\
$^{2}$Institute for Materials Research, Tohoku University, Sendai 980-8577, Japan.\\
$^{3}$Department of Applied Physics, The University of Tokyo, Bunkyo-ku, Tokyo 113-8656, Japan.\\
$^{4}$Advanced Science Research Center, Japan Atomic Energy Agency, Tokai 319-1195, Japan.\\
$^{5}$Advanced Institute for Materials Research, Tohoku University, Sendai 980-8577, Japan.}

\date{\today}

\begin{abstract}
Chirality is breaking of mirror symmetry in matter. In the fields of biology and chemistry, this is particularly important because some of the essential molecules in life such as amino acids and DNA have chirality. It is a long-standing mystery how one of the enantiomers was chosen at the beginning stage of life\cite{chiral1,chiral2}. The understanding of the emergence of homochirality under some conditions is indispensable for the study of the origin of life as well as pharmaceutical science. The chirality is also emergent in magnetic structures. The longitudinal helical magnetic structure is the chiral object composed of magnetic moments, in which the ordered direction of the magnetic moment spatially rotates in the plane perpendicular to the propagation vector (Fig. 1a). Since the sense of rotation, which is denoted as helicity, is reversed by any mirror operation, it is corresponding to the chirality. Here we show that the chirality of a longitudinal helical structure can be controlled by the magnetic field and electric current owing to the spin-transfer torque irrelevant to the spin-orbit interaction and probed by electrical magnetochiral effect, which is sensitive to the chiral symmetry breaking, in an itinerant helimagnet MnP. This phenomenon is distinct from the multiferroicity in transverse-type insulating helical magnets\cite{Fiebig,cheong,tokura,STT}, in which the helical plane is parallel to the propagation vector, because the magnetic structure has polar symmetry not chiral one. While the combination of the magnetic field and electric current satisfies the symmetrical rule of external stimulus for the chirality control\cite{chiral1,chiral2}, the control with them was not reported for any chiral object previously. The present result may pave a new route to the control of chiralities originating from magnetic and atomical arrangements.
\end{abstract}

\maketitle

There are several origins of the helical magnetic structure. One of them is noncentrosymmetric crystal structure. The breaking of space inversion symmetry induces the antisymmetric magnetic interaction denoted as the Dzyaloshinskii-Moriya (DM) interaction, which gives rise to the long period helical spin structure\cite{ishikawa,nakanishi}. Recently this class of helimagnets has been studied extensively because topological spin structures such as skyrmions are frequently observed\cite{skyrmion1,skyrmion2}. In this case, the helicity is locked to the noncentrosymmetric crystal structure and cannot be reversed by the external fields. Another origin of the helical magnetic structure is magnetic frustration. Competing magnetic interactions give rise to helical magnetic structures\cite{yoshimori}. In itinerant systems, the magnetic interactions mediated by the conduction electrons such as RKKY interaction and nesting of Fermi surface also give rise to the helical magnetic structure\cite{koehler,fretwell}. In the latter two mechanisms, the helical magnetic transition involves the breaking of space-inversion symmetry and the spin helicity works as an internal degree of freedom.

\begin{figure}
\centering
\includegraphics[width=14cm]{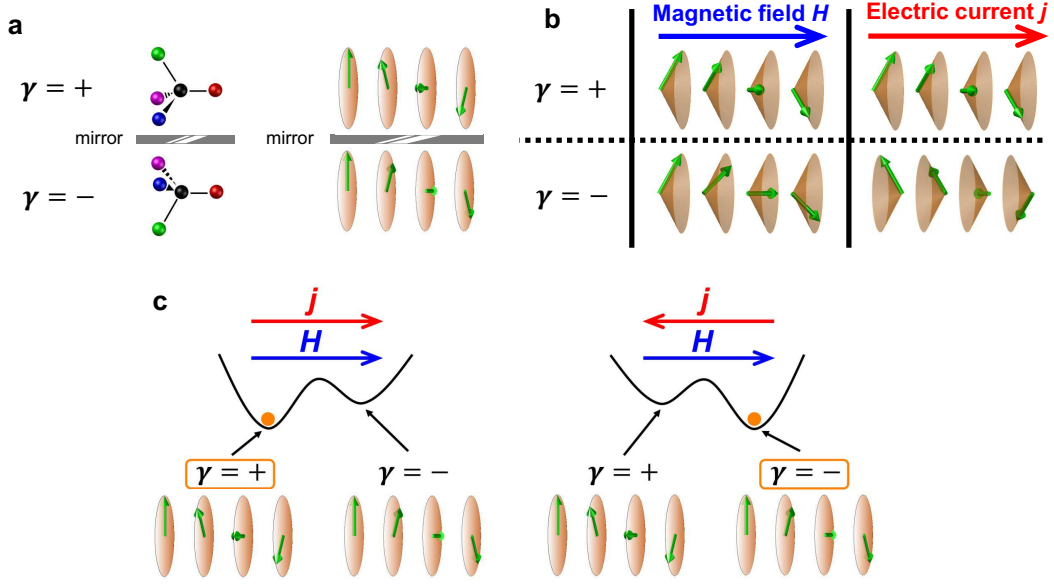}
\caption{\textbf{Electric current control of  helicity in an itinerant longitudinal helimagnet.} \textbf{a} Illustrations of chiralities of molecules and helimagnets. $\gamma =\pm$ stands for the chirality. \textbf{b} Illustrations of the magnetic structure of helimagnets in the presence of a magnetic field or electric current. In the presence of a magnetic field, the magnetic moments are canted to the field direction, forming a conical magnetic structure. The net magnetization is along the magnetic field irrespective of the chirality. The electric current along the helical wave vector also induces the conical magnetic structure. In this case, the net magnetization direction is dependent on the chirality. \textbf{c} Illustrations of the electric current control of helicity in a longitudinal helical magnet. The energetically favored helicity in the presence of magnetic field $H$ and electric current $j$ depends on whether they are parallel or antiparallel to each other.}
\end{figure}

It is of great interest how the spin helicity is coupled to the electromagnetic fields in the latter classes of helimagnets. In the case of an insulator, the electromagnetic coupling has been extensively studied in the field of multiferroics\cite{Fiebig,cheong,tokura}. The ferroelectric polarization is induced by the helical magnetic structure. The sign of polarization is determined by the spin helicity, which can be controlled by the electric field. The magnitude of ferroelectric polarization depends on the angle between the helical plane and the propagation vector; the polarization is maximized when the propagation vector is parallel to the helical plane (transverse helical structure) and vanished for the perpendicular configuration (longitudinal helical structure). These are related to the magnetic symmetry. In the transverse helical structure, the helicity is unchanged by the mirror operation parallel to the helical plane and by that perpendicular to the propagation vector but reversed by the mirror perpendicular to both the former two mirrors, being consistent with the polar symmetry (see extended data Fig. 1). On the other hand, in the longitudinal helical structure, the helicity is reversed by any mirror operation, which indicates the chiral symmetry.

In itinerant helimagnets, the coupling between spin helicity and electric current has scarcely been studied. It should be noted that the electric field is minimal in metallic media. To elucidate this issue, a theory paper regarding the spin dynamics in longitudinal helical magnets under electric current seems quite helpful\cite{STT}. It suggests that the spin-transfer torque induced by the electric current cants the magnetic moments and the cone-like (conical) spin structure is induced. The net magnetization direction of the conical spin structure depends on the sign of the product of spin helicity and electric current (Fig. 1b). This is in contrast with the effect of the magnetic field; the magnetization direction is independent of helicity. One can easily extrapolate from them that the spin helicity degeneracy is lifted under the magnetic field and the electric current (Fig. 1c). In other words, the spin helicity can be controlled by the magnetic field $H$ and the electric current $j$. Based on this conjecture, we here demonstrate the helicity control with the use of the magnetic field and the electric current for an itinerant helical magnet MnP.

\begin{figure}
\centering
\includegraphics[width=10cm]{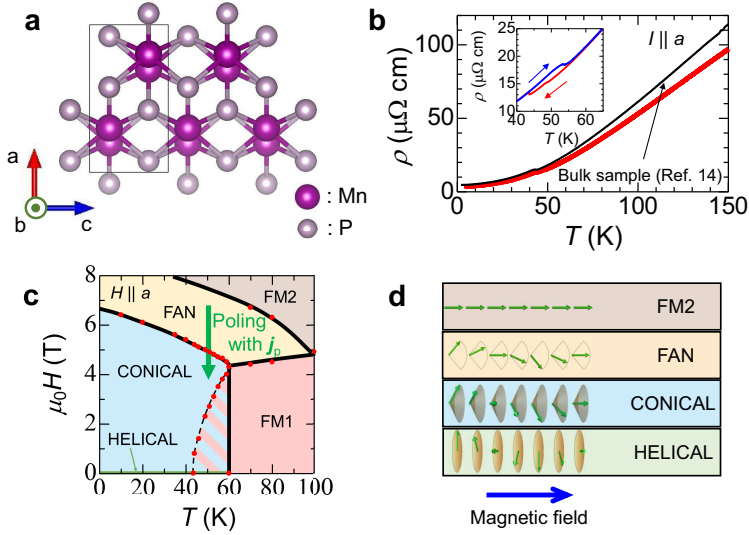}
\caption{\textbf{Properties of the MnP sample.} \textbf{a} Crystal structure of MnP. The black line represents the unit cell. \textbf{b} Temperature dependence of the resistivity for the present MnP sample fabricated with the use of a focused ion beam. The resistivity of bulk (millimeter scale) sample previously measured by Shiomi $et$ $al.$\cite{MnP6} is reproduced for comparison. The inset shows the history dependence of resistivity around the helical-ferromagnetic phase boundary. \textbf{c} The phase diagram of the micro-fabricated MnP sample for $H || a$. The red dots are the phase boundaries estimated by the magnetic field dependence of electrical resistivity (see Extended Data Fig. 3). The black lines are guides for the eyes. In the hatched region, the realized magnetic structure depends on the hysteresis. The green arrow suggests the poling procedure. \textbf{d} Illustrations of the magnetic structures of MnP under magnetic field parallel to the propagation vector.} 
\end{figure}

MnP has orthorhombic and centrosymmetric crystal structure with the space group $Pbnm$ (Fig. 2a)\cite{MnP1,MnP2,MnP3,MnP4,MnP5,MnP6,MnP7,MnPphase}. To increase the electric current density, we fabricated a micron-scale single-crystalline sample (approximately $10\times 20\times 1$ $\rm \mu m^{3}$) with the use of focused ion beam technique (see methods and Extended Data Fig. 2). Figure 2b shows the resistivity for the present micro-fabricated sample compared with that for a bulk (millimeter scale) sample previously reported by Shiomi $et$ $al$\cite{MnP6}. The resistivities are similar to each other, ensuring the sample damage due to the micro-fabrication is minimal for the present sample. In Fig. 2c, we show the magnetic phase diagram constructed based on the magnetic field dependence of electrical resistivity for the present sample (see Extended Data Fig. 3). This is quite similar to that reported in the literature\cite{MnPphase}. The ferromagnetically ordered phase (FM1) emerges below 290 K, in which the magnetic moments are aligned along the $c$-axis. We observed the discontinuous change of resistivity accompanying the hysteresis specific to the first-order phase transition between 40K and 60K (inset of Fig. 2b), indicating the transition to the helical magnetic state. In the helical magnetic phase, the propagation vector is along $a$-axis, and the helical plane is perpendicular to this axis\cite{MnP4}. Therefore, this is the longitudinal helical structure realized in a centrosymmetric material. When the magnetic field is applied parallel to the $a$-axis, the magnetic moments are canted, and a conical magnetic state is realized (Fig. 2d). As the magnetic field increases further before the magnetic moments show the full polarization along to the $a$-axis (FM2), the magnetic structure is transformed into the fan structure, in which the magnetic moments are within $ac$ plane and the angle between the magnetic moment and  the propagation vector spatially oscillates along the propagation vector (Fig. 2d). This magnetic structure can be viewed as the superposition of two conical structures with different helicities. Therefore, the fan-conical magnetic transition is corresponding to the achiral-chiral transition. 

\begin{figure}
\centering
\includegraphics[width=12cm]{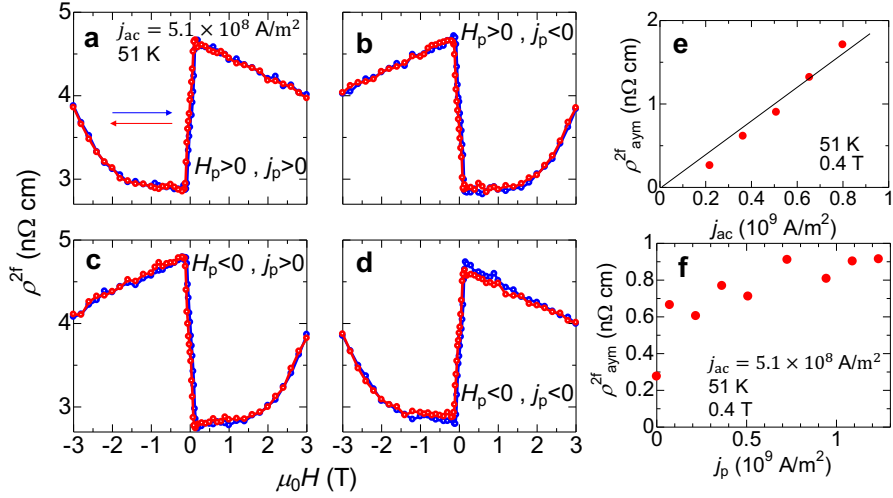}
\caption{\textbf{Electric current control of helicity.} \textbf{a-d} Magnetic field dependence of the 2nd harmonic contribution  of electrical resistivity $\rho^{\rm 2f}$ after the poling procedure with the positive and negative magnetic fields  $H_{\rm p}$ and dc electric currents $j_{\rm p}$. The magnitudes of $j_{\rm p}$ and ac electric current for the measurement $j_{\rm ac}$ are $1.1 \times 10^{9}$ $\rm A/m^{2}$ and $5.1 \times 10^{8}$ $\rm A/m^{2}$, respectively. \textbf{e} $\rho^{\rm 2f}_{\rm asym}$ at 51 K and 0.4 T as a function of $j_{\rm ac}$. \textbf{f} $\rho^{\rm 2f}_{\rm asym}$ at 51 K and 0.4 T as a function of $j_{\rm p}$.}
\end{figure}

In order to control the helicity, we applied the dc electric current $j_{\rm p}$ parallel or antiparallel to the magnetic field $H_{\rm p}$ along to helical wave vector ($a$-axis) when the fan-conical transition field is traversed (Fig. 2c). The magnitude of $H_{\rm p}$ is slowly decreased from 7 T to 3 T at a rate of 4 T/h at 51 K just below the ferromagnetic-helical transition temperature. Then, the dc electric current $j_{\rm p}$ is omitted. This is denoted as the poling procedure. For the detection of controlled helicity after the poling procedure, we utilize the electrical magnetochiral effect\cite{EMC0}. The electrical magnetochiral effect is an electronic transport version of magnetochiral dichroism\cite{mcd}, in which the optical absorption is dependent on the direction of the wave vector irrespective of the polarization.  In the electronic transport phenomenon, the resistivity is dependent on the current as
\begin{equation}
 \rho(\bm{j}) = \rho_{\rm 0} + \rho_{\rm ch}j,
\end{equation}
in which the $\rho_{\rm ch}$ is reversed upon either time-reversal or chirality reversal. In other words, an ac current $j=j_{\rm ac} \sin(\omega t)$ gives rise to the nonlinear voltage
\begin{equation}
 V^{\rm 2nd} (t)=\rho_{\rm ch} j_{\rm ac}^2 \cos (2\omega t). 
\end{equation}
This phenomenon has been observed in several chiral helimagnets\cite{EMC1,EMC2}. Since it has been well established that $\rho_{\rm ch}$ depends on the chirality, it is useful for the probe.  

We show in Figs. 3a-d the 2nd harmonic contribution of electrical resistivity $\rho^{\rm 2f}$ at 51 K as a function of the magnetic field along the helical wave vector ($a$-axis), observed after the poling with $\pm j_{\rm p}$ and $\pm H_{\rm p}$. While the magnetic structure is history-dependent in the low field region at 51 K, the helimagnetic state is stabilized in the field-decreasing process (see Extended Data Fig. 3). The magnitudes of  poling and ac current densities $j_{\rm p}$, $j_{\rm ac}$ are $1.1 \times 10^{9}$ $\rm A/m^{2}$ and $5.1 \times 10^{8}$ $\rm A/m^{2}$, respectively.
The steep change of $\rho^{\rm 2f}$ is observed around $H$=0 indicating the field-odd contribution. This is a hallmark of electrical magnetochiral phenomena specific to chiral symmetry while the field-even contribution seems to be inevitable effects of nonuniformity and/or electrode geometry as discussed in literatures\cite{EMC2,ideue}. Since the crystal of MnP is centrosymmetric, the emergence of magnetochiral phenomenon should be caused by the helical magnetic order. Importantly, the magnetochiral contribution shows a sign change by the reversal of either $H_{\rm p}$ or $j_{\rm p}$. These clearly show that the helicity of the helimagnetic structure depends on whether $H_{\rm p}$ and $j_{\rm p}$ are parallel or antiparallel.

In order to further examine the electrical magnetochiral effect and helicity control, we extract the intrinsic field-odd component of $\rho^{\rm 2f}$ by the anti-symmetrization of the magnetic field dependence of $\rho^{\rm 2f}$ as $\rho^{\rm 2f}_{\rm asym} = (\rho^{\rm 2f} (H$) $-$ $\rho^{\rm 2f} (-H$))/2. As shown in Fig. 3e, $\rho^{\rm 2f}_{\rm asym}$ at 0.4 T is proportional to $j_{\rm ac}$, being consistent with the magnetochiral origin. In Fig. 3f, we show $\rho^{\rm 2f}_{\rm asym}$ at 51 K and 0.4 T as a function of $j_{\rm p}$. $\rho^{\rm 2f}_{\rm asym}$ is saturated above $j_{\rm p} = 1 \times 10^{9}$ $\rm A/m^{2}$, suggesting that the volume fraction of controlled helicity is nearly unity.

\begin{figure}
\centering
\includegraphics[width=5cm]{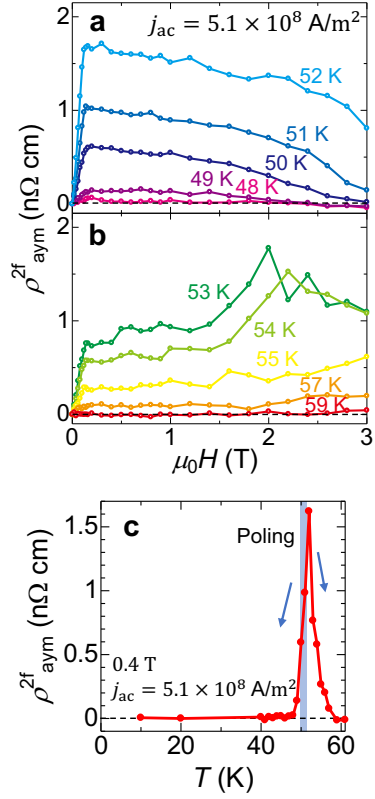}
\caption{\textbf{Temperature and magnetic field dependences of magnetochiral effect.} \textbf{a,b} Magnetic field dependence of $\rho^{\rm 2f}_{\rm asym}$ at various temperatures measured after poling procedure at 51 K with $j_{\rm p} = 1.1 \times 10^{9}$ $\rm A/m^{2}$. \textbf{c} Temperature dependence of $\rho^{\rm 2f}_{\rm asym}$ at 0.4 T.}
\end{figure}

The chirality in this system is totally induced by the magnetic structure. This seems to be reflected by the magnetic field and temperature dependences of $\rho^{\rm 2f}_{\rm asym}$. In Figs. 4a, b, we show the magnetic field dependence of $\rho^{\rm 2f}_{\rm asym}$ below 3 T at various temperatures measured after the poling procedure at 51 K. At 51 K, $\rho^{\rm 2f}_{\rm asym}$ increases steeply and is saturated at a low magnetic field. This magnetic field dependence is contrastive with the linear magnetization curve along $a$-axis\cite{MnP1}. When the temperature is decreased from 51 K, $\rho^{\rm 2f}_{\rm asym}$ steeply decreases and is almost vanished below 48 K. On the other hand, when the temperature is increased from 51 K, $\rho^{\rm 2f}_{\rm asym}$ rapidly increases up to 52 K and then decreases with temperature. Figure 4c shows the temperature dependence of $\rho^{\rm 2f}_{\rm asym}$ at 0.4 T. This clearly shows the sharp enhancement of $\rho^{\rm 2f}_{\rm asym}$ at the phase boundary. In the helimagnets with noncentrosymmetric crystal structures, the magnetochiral effect has been observed in the wider temperature region\cite{EMC1,EMC2}. In MnSi, the magnetochiral signal is enhanced around the helical-paramagnetic transition temperature and suppressed in the high magnetic field\cite{EMC1}. The origin is ascribed to the chiral magnetic fluctuation in the vicinity of the phase boundary. The enhancement of magnetochiral signal around the phase boundary is also observed in $\rm CrNb_{3}S_{6}$ while the overlap of signals with different origins is suggested\cite{EMC2}. In the present case, the magnetochiral signal is more sharply enhanced when the temperature is increased toward the ferromagnetic phase transition temperature. Then, it rapidly decreases above 52 K owing to the reduction of helical volume fraction because the inversion symmetry is unbroken in the ferromagnetic state.  While the transition to the induced ferromagnetic field is observed at low temperature in the DM-induced helimagnets, the helical or conical state is stable in the wide range of magnetic field in the present system.
The robustness of chiral fluctuation against the magnetic field may explain the magnetic field dependence of $\rho^{\rm 2f}_{\rm asym}$. Importantly, the strong chiral fluctuation in the vicinity of phase boundary should make the magnetic structure electrically susceptible, which seems to support the controllability of spin helicity.    

Thus, we have demonstrated the control of helicity  (chirality) with the use of electric current and magnetic field in a longitudinal helical magnet MnP. In the light of symmetry, the chirality can be controlled by collinear polar and axial vectorial fields with either both time-reversal odd or both even symmetry\cite{chiral1,chiral2}. While the chirality control with the use of light beam with certain propagation vector and magnetic field was reported recently\cite{chiral3}, that with a magnetic field and any current, which satisfies the symmetrical rule, was not demonstrated previously. The applicability to a wide range of magnetic and chemical chiral objects should be examined.

\section*{Methods}
A single crystal of MnP was prepared by a Bridgman method. A silica tube with Mn and P was placed in a Bridgman furnace and heated to 1200 $^\circ$C in 14 days and kept for 4 days. The tube was moved toward the low-temperature region in temperature gradient ($\sim$ 1-5 $^\circ$C/mm) at the rate of 0.5 mm per hour for 21 days. Microscale thin rectangular piece with the approximate size of $10\times 20\times 1$ $\rm \mu m^{3}$ was extracted from the crystal with use of focused ion beam (FIB) technique. The thin plate was mounted on a silicon wafer and fixed by the FIB-assisted carbon deposition. Gold electrodes for four-probe measurements were fabricated using electron beam lithography and electron beam deposition. We measured the 1st and 2nd harmonic ac resistivities with the electric current frequency of 14.3 Hz in a superconducting magnet.

\section*{acknowledgements}
The crystal growth was carried out by the joint research in the Institute for Solid State Physics, the University of Tokyo with the help of R. Ishii and Z. Hiroi.
The fabrication of the sample device was partly carried by the collaborative research in the Cooperative Research and Development Center for Advanced Materials, Institute of Materials Research, Tohoku University with help of K. Takanashi and T. Seki.
The authors thank G.E.W. Bauer, Y. Shimamoto, Y. Togawa, and J. Ohe for fruitful discussions. This work was in part supported by JSPS KAKENHI Grant Numbers JP16H04008, JP17H05176, JP18K13494, and JP19H05600 and JST ERATO Spin Quantum Rectification Project (JPMJER1402). N. J. is supported by JPSJ fellows (No.19J11151).

\section*{Author contributuions}
N. J. carried out the crystal growth, device fabrication with focused ion beam and electron beam lithography, and measurements of magnetochiral effect. Y. N. contributed to the crystal growth and measurements of magnetochiral effect. H. A. and E. S. contributed to the device fabrication with a focused ion beam. Y. O. conceived and supervised the project. N. J. and Y. O. wrote the paper through the discussion and assistance from Y. N., H. A., and E. S. 

\section*{Competing interests}
The authers declare no competing interests.

\bibliographystyle{naturemag}
\bibliography{MnPref.bib}

\newpage
\clearpage

\section*{EXTENDED DATA}
\renewcommand{\figurename}{Extended Data Fig.}
\setcounter{figure}{0}

\begin{figure}[h]
\centering
\includegraphics[width=12cm]{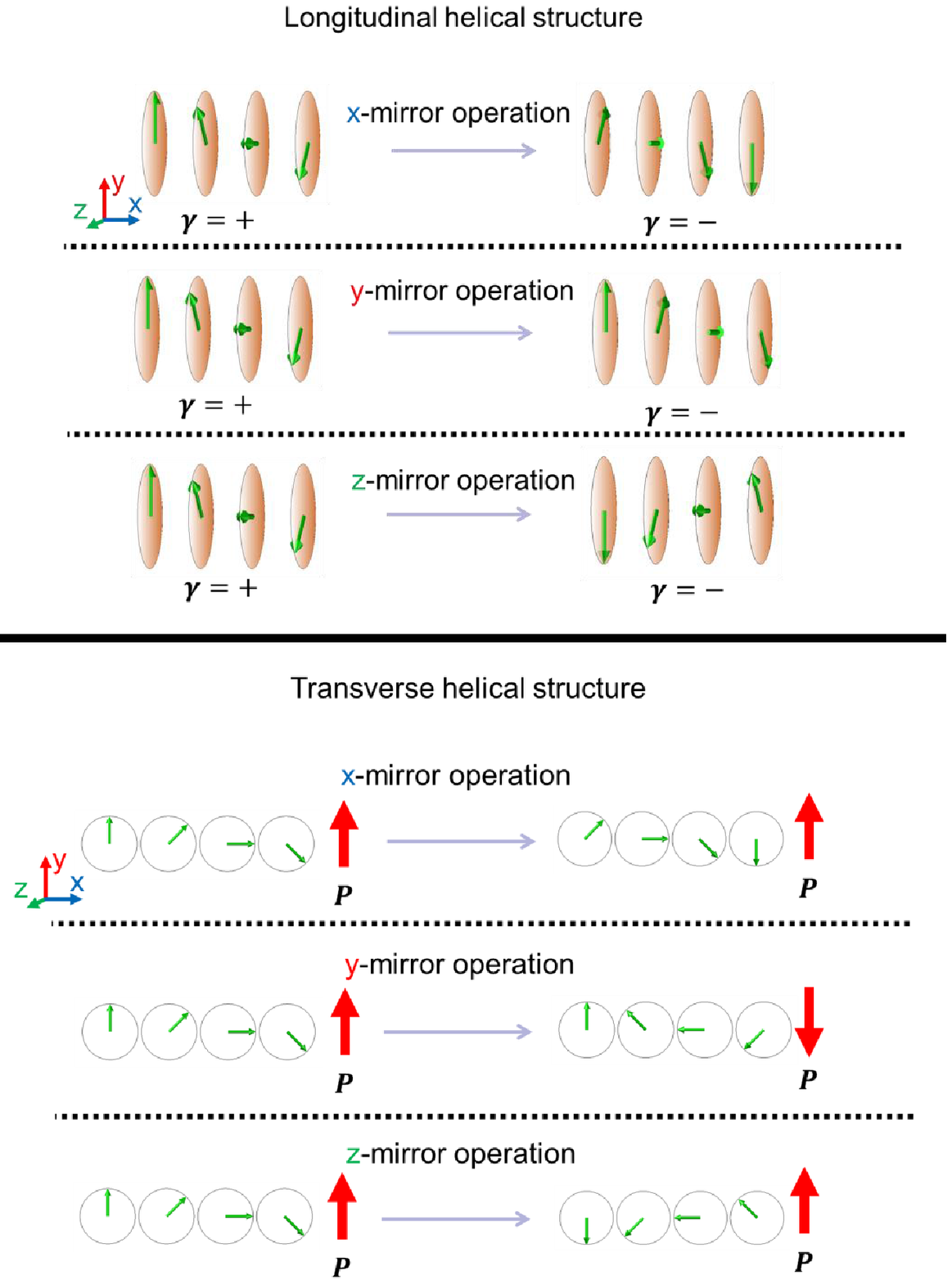}
\caption{\textbf{The effect of mirror operations on longitudinal and transverse helical structures.} $\gamma$ and $P$ stand for the chirality and electric polarization, respectively. x-mirror, y-mirror and z-mirror are the mirrors parallel to x, y, z planes, respectively. } 
\end{figure}

\begin{figure}
\centering
\includegraphics[width=5cm]{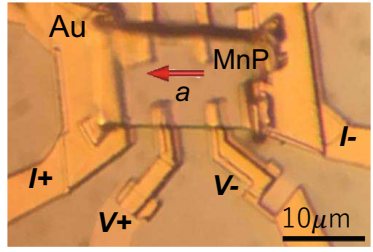}
\caption{\textbf{An optical microscope image of the micro-fabricated MnP sample.} } 
\end{figure}

\begin{figure}
\centering
\includegraphics[width=12cm]{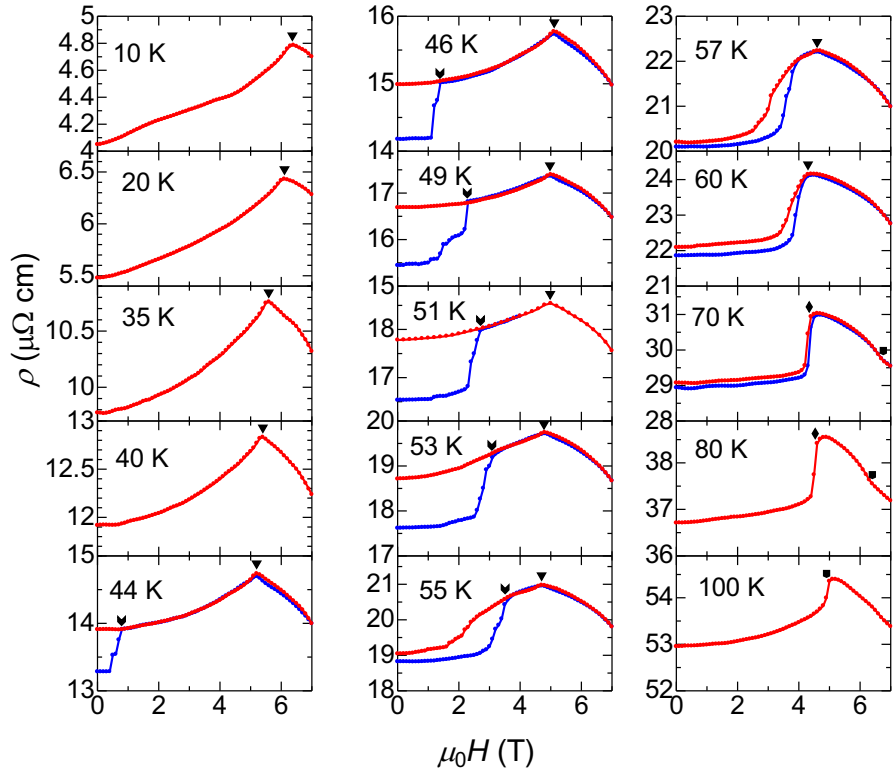}
\caption{\textbf{Magnetic field dependence of the 1st harmonic contribution of electrical resistivity at various temperatures.} The hysteresis is measured between 44 K and 70 K. Before the measurement in this temperature region, the temperature is increased to 80 K and then decreased to the measured temperature in the absence of a magnetic field. The blue data shows the resistivity in the field-increase process just after the zero-field cool. The red curves are measured in the field-decrease process. Finite hysteresis is observed even above 60 K, in which the volume fraction of the helical magnetic phase is thought to vanish. Similar anomalous hysteresis behavior in the FM1 phase was reported in a literature\cite{MnP2}.} 
\end{figure}

\end{document}